\documentclass[11pt]{article}
\usepackage{graphicx}
\usepackage{amsmath}
\usepackage{amssymb}
\usepackage{caption2}
\begin{document}
\bibliographystyle{prsty}
\begin{center}
{\large {\bf \sc{  Analysis of  the ${1\over 2}^{\pm}$ flavor
antitriplet
heavy baryon states   with QCD sum rules }}} \\[2mm]
Zhi-Gang Wang \footnote{E-mail,wangzgyiti@yahoo.com.cn.  }     \\
 Department of Physics, North China Electric Power University,
Baoding 071003, P. R. China

\end{center}

\begin{abstract}
In this article, we study the masses and pole residues of the
${1\over 2}^\pm$ flavor antitriplet heavy baryon  states
($\Lambda_c^+$, $\Xi_c^+,\Xi_c^0)$ and ($\Lambda_b^0$,
$\Xi_b^0,\Xi_b^-)$ by subtracting the contributions from the
corresponding ${1\over 2}^\mp$ heavy baryon  states with the QCD sum
rules, and observe the masses are in good agreement with the
experimental data and make reasonable predictions for the unobserved
${1\over 2}^-$ bottom baryon states. Once reasonable values of the
pole residues $\lambda_{\Lambda}$ and $\lambda_{\Xi}$ are obtained,
we can take them as basic parameters to study the revelent hadronic
processes with the  QCD sum rules.
\end{abstract}

 PACS number: 14.20.Lq, 14.20.Mr

Key words: Heavy baryon states, QCD sum rules

\section{Introduction}

In recent years, several new excited charmed baryon states have been
observed by the BaBar, Belle and CLEO Collaborations, such as the
$\Lambda_c(2765)^+$, $\Lambda_c^+(2880)$, $\Lambda_c^+(2940)$,
$\Sigma_c^+(2800)$, $\Xi_c^+(2980)$, $\Xi_c^+(3077)$,
  $\Xi_c^0(2980)$, $\Xi_c^0(3077)$
  \cite{ShortRV1,ShortRV2,ShortRV3}, and re-vivified  the interest in the
  charmed   baryon spectrum. On the other hand, the QCD sum rules is a powerful
  theoretical tool in studying the
ground state heavy baryon states \cite{SVZ79,PRT85}.  In the QCD sum
rules, the operator product expansion is used to expand the
time-ordered currents into a series of quark and gluon condensates
which parameterize the long distance properties of the QCD vacuum.
Based on the quark-hadron duality, we can obtain copious information
about the hadronic parameters at the phenomenological side
\cite{SVZ79,PRT85}. There have been several works on the masses of
the heavy baryon states with the full QCD sum rules and the QCD sum
rules in the heavy quark effective theory
  \cite{Bagan93,Bagan922,Bagan921,M-Nielsen07,Huang0805,M-Huang0811,Narison0904,Shuryak82,Grozin92,
  Dai961,Dai962,Huang02,Zhu00,HuangCS,HuangMQ,M-Liu07}.

In Refs.\cite{Wang0704,Wang0809,Wang0910}, we study the
${\frac{1}{2}}^+$ heavy baryon states $\Omega_Q$, $\Xi'_Q$ and
$\Sigma_Q$ and ${\frac{3}{2}}^+$ heavy baryon states $\Omega_Q^*$,
$\Xi^*_Q$ and $\Sigma^*_Q$ with the full QCD sum rules, and observe
that the pole residues of the ${\frac{3}{2}}^+$ heavy baryon states
from the sum rules with different tensor structures are consistent
with each other, while the pole residues of the ${\frac{1}{2}}^+$
heavy baryon states from the sum rules with different tensor
structures differ from each other greatly. Those pole residues are
important parameters in studying the radiative decays $\Omega_Q^*\to
\Omega_Q \gamma$, $\Xi_Q^*\to \Xi'_Q \gamma$ and $\Sigma_Q^*\to
\Sigma_Q \gamma$ \cite{Wang0910,Wang0909}, we should refine those
parameters to improve the predictive ability.

In Ref.\cite{Oka96}, Jido et al introduce a novel approach based on
the QCD sum rules to separate the contributions of   the
negative-parity light flavor  baryon states from the positive-parity
light flavor baryon states, as the interpolating currents may have
non-vanishing couplings to both the negative- and positive-parity
baryon states \cite{Chung82}. Before the work of Jido et al, Bagan
et al take the infinite mass limit for the heavy quarks to separate
the contributions of the positive and negative parity heavy baryon
states unambiguously \cite{Bagan93}.

In Ref.\cite{Wang0912}, we follow Ref.\cite{Oka96} and re-study the
masses and pole residues of the ${\frac{1}{2}}^+$ flavor sextet
heavy baryon states $\Omega_Q$, $\Xi'_Q$ and $\Sigma_Q$ by
subtracting the contributions of  the negative parity heavy baryon
states with the full QCD sum rules. In this article, we use the same
approach to study the ${1\over 2}^{\pm}$ flavor antitriplet heavy
baryon states ($\Lambda_c^+$, $\Xi_c^+,\Xi_c^0)$ and ($\Lambda_b^0$,
$\Xi_b^0,\Xi_b^-)$.

The article is arranged as follows:  we derive the QCD sum rules for
the masses and the pole residues of  the heavy baryon states
($\Lambda_c^+$, $\Xi_c^+,\Xi_c^0)$ and ($\Lambda_b^0$,
$\Xi_b^0,\Xi_b^-)$  in Sect.2; in Sect.3, we present the
 numerical results and discussions; and Sect.4 is reserved for our
conclusions.

\section{QCD sum rules for  the $\Lambda_Q$ and $\Xi_Q$ }
The ${1\over 2}^+$ flavor  antitriplet heavy baryon states
($\Lambda_c^+$, $\Xi_c^+,\Xi_c^0)$ and ($\Lambda_b^0$,
$\Xi_b^0,\Xi_b^-)$ can be interpolated by the following currents
$J_\Lambda(x)$ and $J_\Xi(x)$
 respectively,
\begin{eqnarray}
J_\Lambda(x)&=& \epsilon^{ijk}  u^T_i(x)C\gamma_5 d_j(x)   Q_k(x)  \, ,  \nonumber \\
J_\Xi(x)&=& \epsilon^{ijk}  q^T_i(x)C\gamma_5 s_j(x)   Q_k(x)  \, ,
\end{eqnarray}
where the  $Q$ represents the heavy quarks $c$ and $b$,  the $i$,
$j$ and $k$ are color indexes, and the $C$ is the charge conjunction
matrix.

 The corresponding negative-parity heavy baryon states can be
interpolated by the  currents $J_{-} =i\gamma_{5} J_{+}$  because
multiplying $i \gamma_{5}$ to the $J_{+}$ changes the parity of the
$J_{+}$ \cite{Oka96}, where the $J_{+}$ denotes the currents
$J_\Lambda(x)$ and $J_\Xi(x)$. The correlation functions are defined
by
\begin{eqnarray}
\Pi_{\pm}(p)&=&i\int d^4x e^{ip \cdot x} \langle
0|T\left\{J_{\pm}(x)\bar{J}_{\pm}(0)\right\}|0\rangle \, ,
\end{eqnarray}
and can be decomposed as
\begin{equation}
    \Pi_{\pm}(p) = \!\not\!{p} \Pi_{1}(p) \pm \Pi_{0}(p)\, ,
\end{equation}
due to Lorentz covariance. The currents $J_{\pm}$ couple  to both
the positive-  and negative-parity baryon states \cite{Chung82},
i.e. $ \langle{0}|J_{\pm}| B^{-}\rangle \langle
B^{-}|\bar{J}_{\pm}|0\rangle =     - \gamma_{5}\langle 0|J_{\mp}|
B^{-}\rangle \langle B^{-}| \bar{J}_{\mp}|0\rangle
    \gamma_{5}$, where the $B^{-}$ denote the negative parity baryon states.

 We  insert  a
complete set  of intermediate baryon states with the same quantum
numbers as the current operators $J_{+}(x)$ and $J_{-}(x)$ into the
correlation functions $\Pi_{\pm}(p)$  to obtain the hadronic
representation \cite{SVZ79,PRT85}. After isolating the pole terms of
the lowest states, we obtain the following result \cite{Oka96}:
\begin{eqnarray}
    \Pi_{\pm}(p)     & = &   \lambda_+^2 {\!\not\!{p} +
    M_{+} \over M^{2}_+ -p^{2} } + \lambda_{-}^2
    {\!\not\!{p} - M_{-} \over M_{-}^{2}-p^{2}  } +\cdots \, ,
    \end{eqnarray}
where the $M_{\pm}$ are the masses of the lowest states with parity
$\pm$ respectively, and the $\lambda_{\pm}$ are the  corresponding
pole residues (or couplings).
 If we take $\vec{p} = 0$, then
\begin{eqnarray}
  \rm{limit}_{\epsilon\rightarrow0}\frac{{\rm Im}  \Pi_{\pm}(p_{0}+i\epsilon)}{\pi} & = &
    \lambda_+^2 {\gamma_{0} + 1\over 2} \delta(p_{0} - M_+) +
    \lambda_{-}^{2} {\gamma_{0} - 1\over 2} \delta(p_{0} - M_{-})+\cdots \nonumber \\
  & = & \gamma_{0} A(p_{0}) \pm B(p_{0})+\cdots \, ,
\end{eqnarray}
where
\begin{eqnarray}
  A(p_{0}) & = & {1 \over 2} \left[ \lambda_+^{2}
  \delta(p_{0} - M_+)  + \lambda_-^{2} \delta(p_{0} -
  M_{-})\right] \, , \nonumber \\
   B(p_{0}) & = & \pm{1 \over 2} \left[ \lambda_+^{2}
  \delta(p_{0} - M_+)  - \lambda_-^{2} \delta(p_{0} -
  M_{-})\right] \, ,
\end{eqnarray}
the contribution $A(p_{0}) + B(p_{0})$ ($A(p_{0}) - B(p_{0})$)
contains contributions  from the positive parity (negative parity)
states only for the $\Pi_+(p_0)$, while the contribution $A(p_{0}) +
B(p_{0})$ ($A(p_{0}) - B(p_{0})$) contains contributions  from the
negative parity (positive parity) states only for the $\Pi_-(p_0)$.

We  calculate the light quark parts of the correlation functions
$\Pi_{\pm}(p)$ in the coordinate space and use the momentum space
expression for the heavy quark propagators, then resort to the
Fourier integral to transform  the light quark parts into the
momentum space in $D$ dimensions,  take $\vec{p} = 0$,  and  use the
dispersion relation to obtain the spectral densities $\rho^A(p_0)$
and $\rho^B(p_0)$ (which correspond to the tensor structures
$\gamma_0$ and $1$ respectively) at the level of quark-gluon degrees
of freedom, finally we introduce the weight functions
$\exp\left[-\frac{p_0^2}{T^2}\right]$,
$p_0^2\exp\left[-\frac{p_0^2}{T^2}\right]$,   and obtain the
following sum rules,
\begin{eqnarray}
  \lambda_{\pm}^2\exp\left[-\frac{M_{\pm}^2}{T^2}\right]&=&\int_{\Delta}^{\sqrt{s_0}}dp_0
\left[\rho^A(p_0)
+\rho^B(p_0)\right]\exp\left[-\frac{p_0^2}{T^2}\right] \, ,
\end{eqnarray}
\begin{eqnarray}
  \lambda_{\pm}^2M_{\pm}^2\exp\left[-\frac{M_{\pm}^2}{T^2}\right]&=&\int_{\Delta}^{\sqrt{s_0}}dp_0
\left[\rho^A(p_0)
+\rho^B(p_0)\right]p_0^2\exp\left[-\frac{p_0^2}{T^2}\right] \, ,
\end{eqnarray}
where
\begin{eqnarray}
\rho^A_{\Xi_Q}(p_0)&=&\frac{3p_0}{128\pi^4}\int_{t_i}^1dt
t(1-t)^2(p_0^2-\widetilde{m}_Q^2)^2+\frac{p_0m_s\left[\langle\bar{s}s\rangle-2\langle\bar{q}q\rangle\right]}{16\pi^2}\int_{t_i}^1
dt t\nonumber\\
&&+\frac{p_0}{128\pi^2}\langle \frac{\alpha_sGG}{\pi}\rangle
\int_{t_i}^1 dt t-\frac{m_Q^2}{768\pi^2}\langle
\frac{\alpha_sGG}{\pi}\rangle \int_{t_i}^1 dt \frac{(1-t)^2}{t^2}
\delta
(p_0-\widetilde{m}_Q)\nonumber \\
&&+\frac{m_s\left[3\langle\bar{q}g_s\sigma
Gq\rangle-\langle\bar{s}g_s\sigma Gs\rangle \right]}{192\pi^2}\delta
(p_0-m_Q) +\frac{
\langle\bar{q}q\rangle\langle\bar{s}s\rangle}{12}\delta(p_0-m_Q) \,
,
\end{eqnarray}

\begin{eqnarray}
\rho^B_{\Xi_Q}(p_0)&=&\frac{3m_Q}{128\pi^4}\int_{t_i}^1dt
(1-t)^2(p_0^2-\widetilde{m}_Q^2)^2+\frac{m_sm_Q\left[\langle\bar{s}s\rangle-2\langle\bar{q}q\rangle\right]}{16\pi^2}\int_{t_i}^1
dt \nonumber\\
&&+\frac{m_Q}{128\pi^2}\langle \frac{\alpha_sGG}{\pi}\rangle
\int_{t_i}^1 dt +\frac{m_Q}{192\pi^2}\langle
\frac{\alpha_sGG}{\pi}\rangle
\int_{t_i}^1 dt \frac{(1-t)^3}{t^2}\nonumber \\
&&-\frac{m_Q}{768\pi^2}\langle \frac{\alpha_sGG}{\pi}\rangle
\int_{t_i}^1 dt \frac{(1-t)^2}{t}\widetilde{m}_Q
 \delta(p_0-\widetilde{m}_Q)\nonumber \\
&&+\frac{m_s\left[3\langle\bar{q}g_s\sigma
Gq\rangle-\langle\bar{s}g_s\sigma Gs\rangle \right]}{192\pi^2}\delta
(p_0-m_Q) +\frac{
\langle\bar{q}q\rangle\langle\bar{s}s\rangle}{12}\delta(p_0-m_Q) \,
,
\end{eqnarray}

\begin{eqnarray}
\rho^A_{\Lambda_Q}(p_0)&=&\frac{3p_0}{128\pi^4}\int_{t_i}^1dt
t(1-t)^2(p_0^2-\widetilde{m}_Q^2)^2+\frac{ \langle\bar{q}q\rangle^2}{12}\delta(p_0-m_Q)\nonumber\\
&&+\frac{p_0}{128\pi^2}\langle \frac{\alpha_sGG}{\pi}\rangle
\int_{t_i}^1 dt t  -\frac{m_Q^2}{768\pi^2}\langle
\frac{\alpha_sGG}{\pi}\rangle \int_{t_i}^1 dt \frac{(1-t)^2}{t^2}
\delta (p_0-\widetilde{m}_Q) \, ,
\end{eqnarray}

\begin{eqnarray}
\rho^B_{\Lambda_Q}(p_0)&=&\frac{3m_Q}{128\pi^4}\int_{t_i}^1dt
(1-t)^2(p_0^2-\widetilde{m}_Q^2)^2 +\frac{
\langle\bar{q}q\rangle^2}{12}\delta(p_0-m_Q) \nonumber\\
&&+\frac{m_Q}{128\pi^2}\langle \frac{\alpha_sGG}{\pi}\rangle
\int_{t_i}^1 dt +\frac{m_Q}{192\pi^2}\langle
\frac{\alpha_sGG}{\pi}\rangle
\int_{t_i}^1 dt \frac{(1-t)^3}{t^2}\nonumber \\
&&-\frac{m_Q}{768\pi^2}\langle \frac{\alpha_sGG}{\pi}\rangle
\int_{t_i}^1 dt \frac{(1-t)^2}{t}\widetilde{m}_Q
 \delta(p_0-\widetilde{m}_Q)  \, ,
\end{eqnarray}
where $\widetilde{m}_Q^2=\frac{m_Q^2}{t}$,
$t_i=\frac{m_Q^2}{p_0^2}$, the $s_0$ are the threshold parameters,
$T^2$ is the Borel parameter,  $\Delta=m_Q+m_s$ and $m_Q$ in the
channels $\Xi_Q$ and $\Lambda_Q$ respectively.

\section{Numerical results and discussions}
The input parameters are taken to be the standard values $\langle
\bar{q}q \rangle=-(0.24\pm 0.01 \,\rm{GeV})^3$,  $\langle \bar{s}s
\rangle=(0.8\pm 0.2 )\langle \bar{q}q \rangle$, $\langle
\bar{q}g_s\sigma Gq \rangle=m_0^2\langle \bar{q}q \rangle$, $\langle
\bar{s}g_s\sigma Gs \rangle=m_0^2\langle \bar{s}s \rangle$,
$m_0^2=(0.8 \pm 0.2)\,\rm{GeV}^2$ \cite{Ioffe2005,LCSRreview},
$\langle \frac{\alpha_s GG}{\pi}\rangle=(0.012 \pm
0.004)\,\rm{GeV}^4 $ \cite{LCSRreview},
$m_s=(0.14\pm0.01)\,\rm{GeV}$, $m_c=(1.35\pm0.10)\,\rm{GeV}$ and
$m_b=(4.7\pm0.1)\,\rm{GeV}$ \cite{PDG} at the energy scale  $\mu=1\,
\rm{GeV}$.

  The value of the gluon condensate $\langle \frac{\alpha_s
GG}{\pi}\rangle $ has been updated from time to time, and changes
greatly \cite{NarisonBook}.
 At the present case, the gluon condensate  makes tiny  contribution,  the updated value $\langle \frac{\alpha_s GG}{\pi}\rangle=(0.023 \pm
0.003)\,\rm{GeV}^4 $ \cite{NarisonBook} and the standard value
$\langle \frac{\alpha_s GG}{\pi}\rangle=(0.012 \pm
0.004)\,\rm{GeV}^4 $ \cite{LCSRreview} lead to a difference less
than $15\,\rm{MeV}$ for the masses.

In the conventional QCD sum rules \cite{SVZ79,PRT85}, there are two
criteria (pole dominance and convergence of the operator product
expansion) for choosing  the Borel parameter $T^2$ and threshold
parameter $s_0$.  We impose the two criteria on the heavy baryon
states to choose the Borel parameter $T^2$ and threshold parameter
$s_0$, the values are shown in Table 1. From Table 1, we can see
that the contribution from the perturbative term  is dominant, the
operator product expansion is convergent certainly. In this article,
we take the contribution from the pole term is larger than $45\%$
($49\%$) for the positive (negative) parity baryon states, the
uncertainty of the threshold parameter is $0.1\,\rm{GeV}$, and the
Borel window is $1\,\rm{GeV}^2$.

In calculation, we  neglect  the contributions from the perturbative
corrections.  Those perturbative corrections can be taken into
account in the leading logarithmic
 approximations through  anomalous dimension factors. After the Borel transform, the effects of those
 corrections are  to multiply each term on the operator product
 expansion side by the factor, $ \left[ \frac{\alpha_s(T^2)}{\alpha_s(\mu^2)}\right]^{2\Gamma_{J}-\Gamma_{\mathcal
 {O}_n}}  $,
 where the $\Gamma_{J}$ is the anomalous dimension of the
 interpolating current $J(x)$ and the $\Gamma_{\mathcal {O}_n}$ is the anomalous dimension of
 the local operator $\mathcal {O}_n(0)$. We carry out the operator product expansion at a special energy
scale $\mu^2=1\,\rm{GeV}^2$, and  set the factor $\left[
\frac{\alpha_s(T^2)}{\alpha_s(\mu^2)}\right]^{2\Gamma_{J}-\Gamma_{\mathcal
{O}_n}}\approx1$, such an approximation maybe result in some scale
dependence  and  weaken the prediction ability. In this article, we
study the flavor antitriplet $\frac{1}{2}^{\pm}$ heavy baryon states
systemically,  the predictions are still robust   as we take the
analogous criteria in those sum rules.

Taking into account all uncertainties  of the relevant  parameters,
we obtain the values of the masses and pole residues of
 the  flavor antitriplet $\frac{1}{2}^{\pm}$ heavy baryon states ($\Lambda_c^+$,
$\Xi_c^+,\Xi_c^0)$ and ($\Lambda_b^0$, $\Xi_b^0,\Xi_b^-)$, which are
shown in Figs.1-2 and Table 2. In this article,  we calculate the
uncertainties $\delta$  with the formula
\begin{eqnarray}
\delta=\sqrt{\sum_i\left(\frac{\partial f}{\partial
x_i}\right)^2\mid_{x_i=\bar{x}_i} (x_i-\bar{x}_i)^2}\,  ,
\end{eqnarray}
 where the $f$ denote  the
hadron mass  $M$ and the pole residue $\lambda$,  the $x_i$ denote
the input QCD parameters $m_c$, $m_b$, $\langle \bar{q}q \rangle$,
$\langle \bar{s}s \rangle$, $\cdots$, and the threshold parameter
$s_0$ and Borel parameter $T^2$. As the partial
 derivatives   $\frac{\partial f}{\partial x_i}$ are difficult to carry
out analytically, we take the  approximation $\left(\frac{\partial
f}{\partial x_i}\right)^2 (x_i-\bar{x}_i)^2\approx
\left[f(\bar{x}_i\pm \Delta x_i)-f(\bar{x}_i)\right]^2$ in the
numerical calculations.

\begin{table}
\begin{center}
\begin{tabular}{|c|c|c|c|c|c|}
\hline\hline & $T^2 (\rm{GeV}^2)$& $\sqrt{s_0} (\rm{GeV})$&pole&perturbative\\
\hline
     $\Lambda_c({\frac{1}{2}}^+)$  &$1.7-2.7$ &$3.1$& $(46-83)\%$ &$(47-73)\%$\\ \hline
        $\Xi_c({\frac{1}{2}}^+)$  &$1.9-2.9$ &$3.2$ & $(46-79)\%$&$(59-77)\%$\\ \hline
   $\Lambda_b({\frac{1}{2}}^+)$  &$4.3-5.3$ &$6.5$&  $(46-67)\%$&$(58-72)\%$\\ \hline
    $\Xi_b({\frac{1}{2}}^+)$  &$4.4-5.4$ &$6.5$& $(45-64)\%$&$(62-73)\%$\\ \hline
  $\Lambda_c({\frac{1}{2}}^-)$  &$2.2-3.2$ &$3.4$& $(49-77)\%$&$(70-84)\%$\\ \hline
 $\Xi_c({\frac{1}{2}}^-)$  &$2.4-3.4$ &$3.5$ & $(49-75)\%$&$(76-86)\%$\\ \hline
 $\Lambda_b({\frac{1}{2}}^-)$  &$4.7-5.7$ &$6.7$& $(49-67)\%$&$(69-80)\%$\\ \hline
 $\Xi_b({\frac{1}{2}}^-)$  &$5.0-6.0$ &$6.8$& $(49-65)\%$&$(75-83)\%$\\ \hline\hline
\end{tabular}
\end{center}
\caption{ The Borel parameters $T^2$ and threshold parameters $s_0$
for the heavy baryon states, the "pole" stands for the contribution
from the pole term, and the "perturbative" stands for the
contribution from the perturbative term in the operator product
expansion.}
\end{table}

\begin{table}
\begin{center}
\begin{tabular}{|c|c|c|c|c|c|c|}
\hline\hline & $T^2 (\rm{GeV}^2)$& $\sqrt{s_0} (\rm{GeV})$&
$M(\rm{GeV})$&$\lambda
(\rm{GeV}^3)$&$M(\rm{GeV})[\rm{exp}]$\cite{PDG}\\\hline
  $\Lambda_c({\frac{1}{2}}^+)$  &$1.7-2.7$ &$3.1\pm0.1$& $2.26\pm0.27$&$0.022\pm0.008$ &2.28646\\ \hline
  $\Xi_c({\frac{1}{2}}^+)$  &$1.9-2.9$ &$3.2\pm0.1$ & $2.44\pm0.23$&$0.027\pm0.008$&2.4678/2.47088\\ \hline
 $\Lambda_b({\frac{1}{2}}^+)$  &$4.3-5.3$ &$6.5\pm0.1$& $5.65\pm0.20$&$0.030\pm0.009$&5.6202\\ \hline
 $\Xi_b({\frac{1}{2}}^+)$  &$4.4-5.4$ &$6.5\pm0.1$& $5.73\pm0.18$&$0.032\pm0.009$&5.7924\\ \hline
 $\Lambda_c({\frac{1}{2}}^-)$  &$2.2-3.2$ &$3.4\pm0.1$& $2.61\pm0.21$&$0.035\pm0.009$ &2.5954\\ \hline
 $\Xi_c({\frac{1}{2}}^-)$  &$2.4-3.4$ &$3.5\pm0.1$ & $2.76\pm0.18$&$0.042\pm0.009$&2.7891/2.7918\\ \hline
 $\Lambda_b({\frac{1}{2}}^-)$  &$4.7-5.7$ &$6.7\pm0.1$&  $5.85\pm0.18$&$0.042\pm0.012$&?\\ \hline
 $\Xi_b({\frac{1}{2}}^-)$  &$5.0-6.0$ &$6.8\pm0.1$& $6.01\pm0.16$&$0.051\pm0.012$&?\\ \hline
      \hline
\end{tabular}
\end{center}
\caption{ The masses $M(\rm{GeV})$ and pole residues
$\lambda(\rm{GeV}^3)$ of the heavy baryon states.}
\end{table}

\begin{table}
\begin{center}
\begin{tabular}{|c|c|c|c|c|}
\hline\hline
 &\cite{M-Nielsen07}&\cite{M-Huang0811}&\cite{M-Liu07}&This work\\ \hline
 $\Lambda_c({\frac{1}{2}}^+)$& &$2.31\pm 0.19$ & $2.271^{+0.067}_{-0.049} $& $2.26\pm0.27$ \\ \hline
  $\Xi_c({\frac{1}{2}}^+)$& $2.5\pm0.2$&$2.48 \pm 0.21$ & $2.432^{+0.079}_{-0.068}$& $2.44\pm0.23$ \\ \hline
  $\Lambda_b({\frac{1}{2}}^+)$& &$5.69 \pm 0.13$ & $5.637^{+0.068}_{-0.056}$& $5.65\pm0.20$ \\ \hline
  $\Xi_b({\frac{1}{2}}^+)$& $5.75\pm0.25$&$5.75 \pm 0.13$ & $5.780^{+0.073}_{-0.068}$& $5.73\pm0.18$ \\ \hline
   $\Lambda_c({\frac{1}{2}}^-)$& &$2.53\pm 0.22$ & & $2.61\pm0.21$ \\ \hline
  $\Xi_c({\frac{1}{2}}^-)$& &$2.65 \pm 0.27$ & & $2.76\pm0.18$ \\ \hline
   $\Lambda_b({\frac{1}{2}}^-)$& &$5.85 \pm 0.15$ &  & $5.85\pm0.18$ \\ \hline
 $\Xi_b({\frac{1}{2}}^-)$& &$5.95 \pm 0.16$ & & $6.01\pm0.16$ \\ \hline
 \hline
\end{tabular}
\end{center}
\caption{ The masses $M(\rm{GeV})$   of the  heavy baryon states
from the QCD sum rules.}
\end{table}

\begin{figure}
 \centering
 \includegraphics[totalheight=4cm,width=5cm]{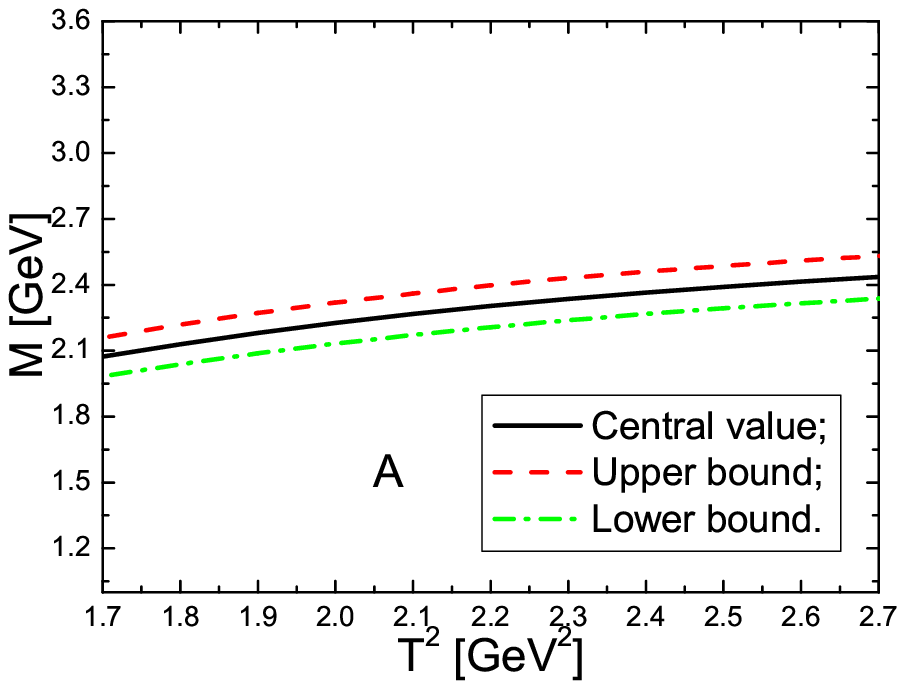}
  \includegraphics[totalheight=4cm,width=5cm]{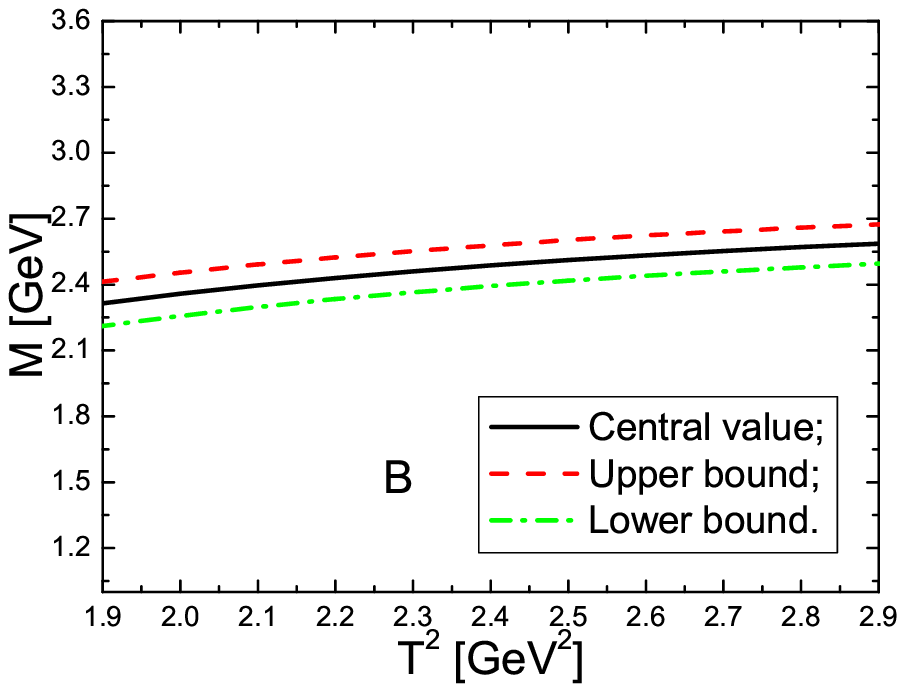}
  \includegraphics[totalheight=4cm,width=5cm]{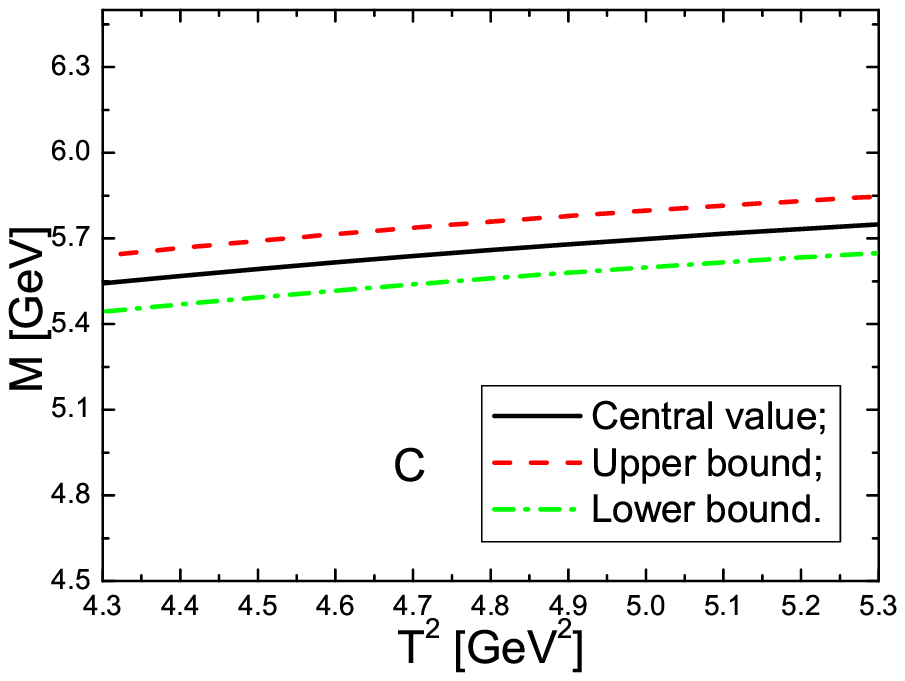}
  \includegraphics[totalheight=4cm,width=5cm]{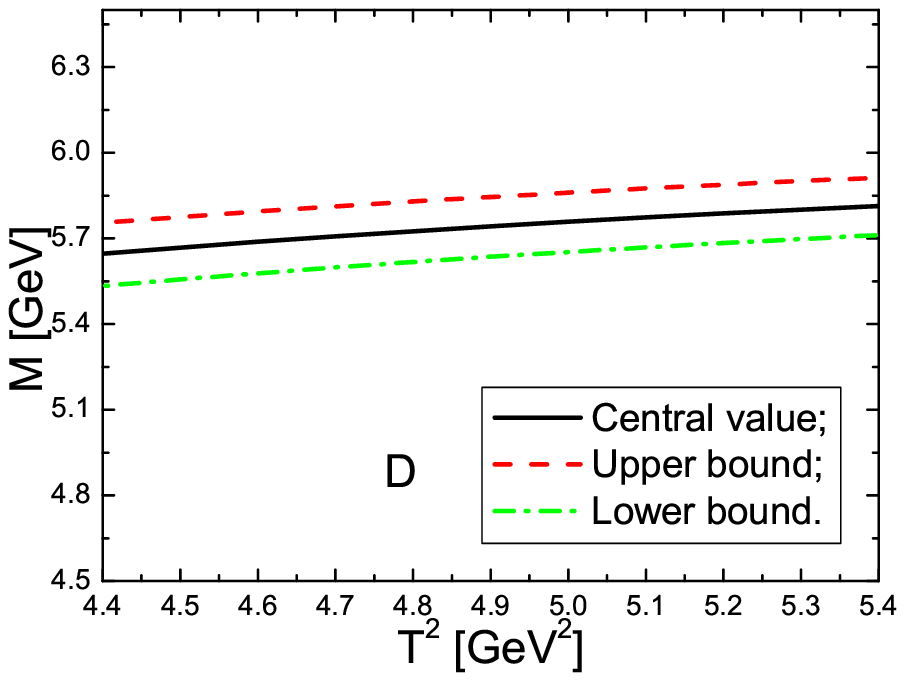}
\includegraphics[totalheight=4cm,width=5cm]{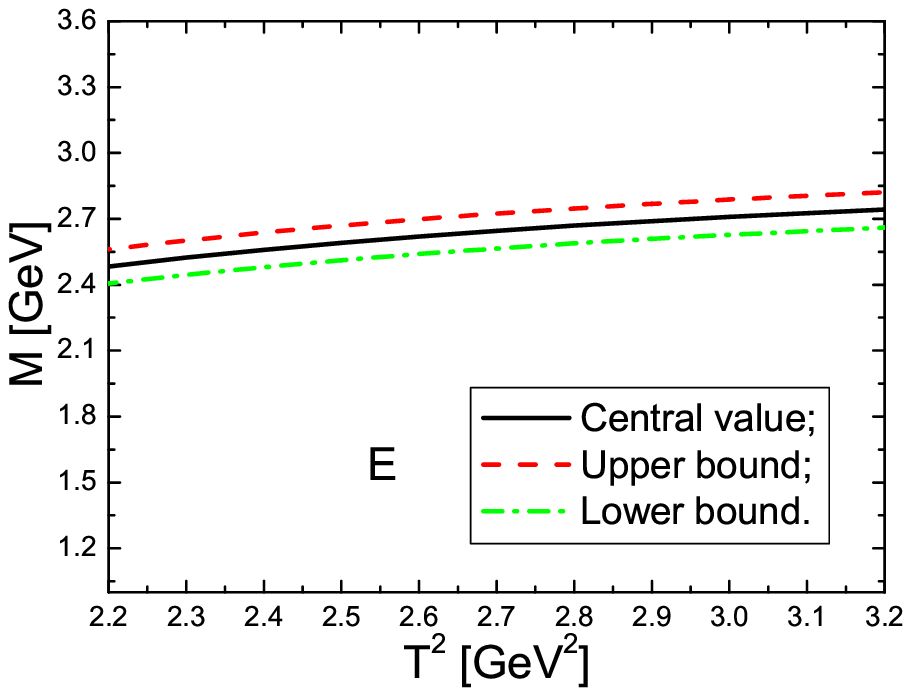}
 \includegraphics[totalheight=4cm,width=5cm]{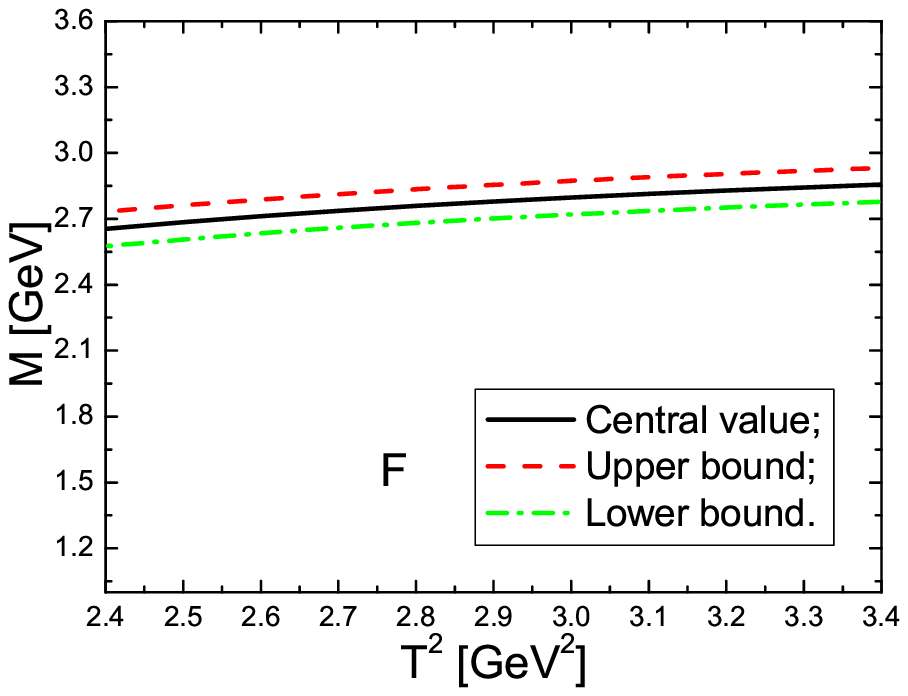}
\includegraphics[totalheight=4cm,width=5cm]{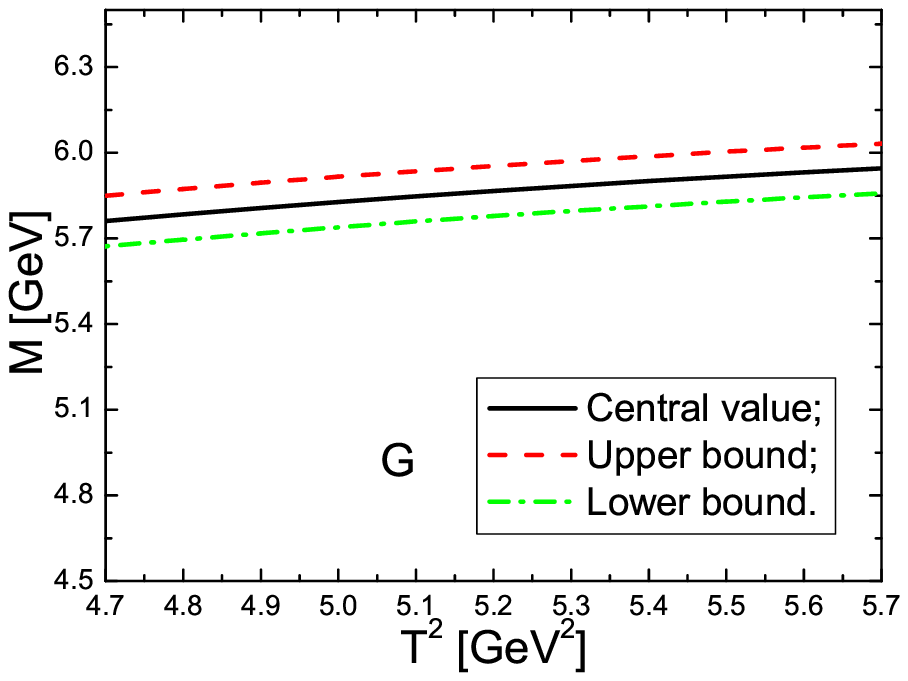}
\includegraphics[totalheight=4cm,width=5cm]{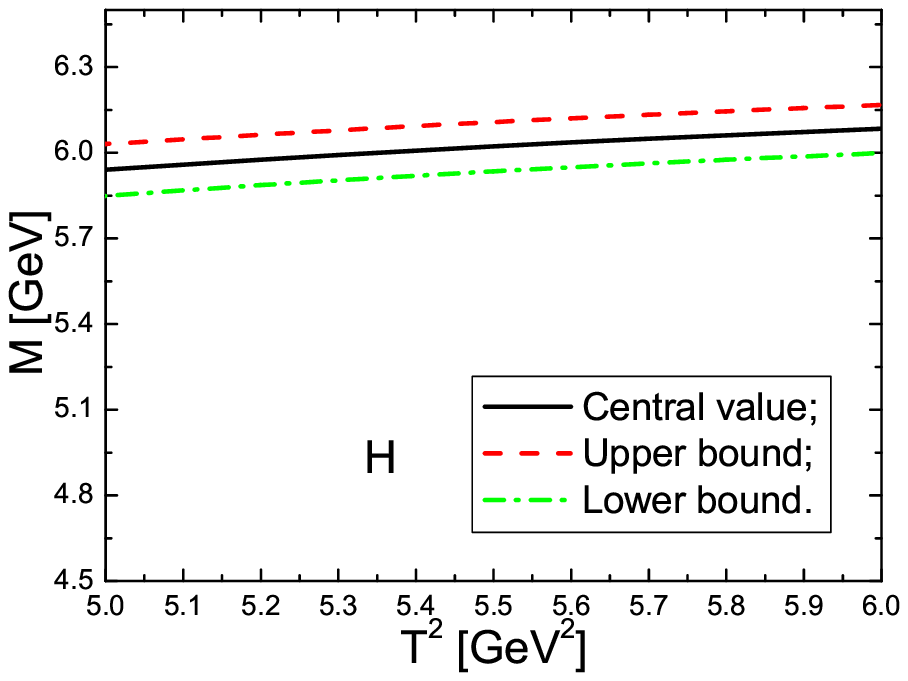}
        \caption{ The masses  $M$ of the heavy baryon states, the $A$, $B$, $C$ and $D$  correspond
       to the ${\frac{1}{2}}^+$ heavy baryon channels $\Lambda_c$, $\Xi_c$, $\Lambda_b$ and $\Xi_b$ respectively, while
the $E$, $F$, $G$ and $H$  correspond        to the
${\frac{1}{2}}^-$ heavy baryon channels $\Lambda_c$, $\Xi_c$,
$\Lambda_b$ and $\Xi_b$ respectively.  }
\end{figure}
\begin{figure}
 \centering
 \includegraphics[totalheight=4cm,width=5cm]{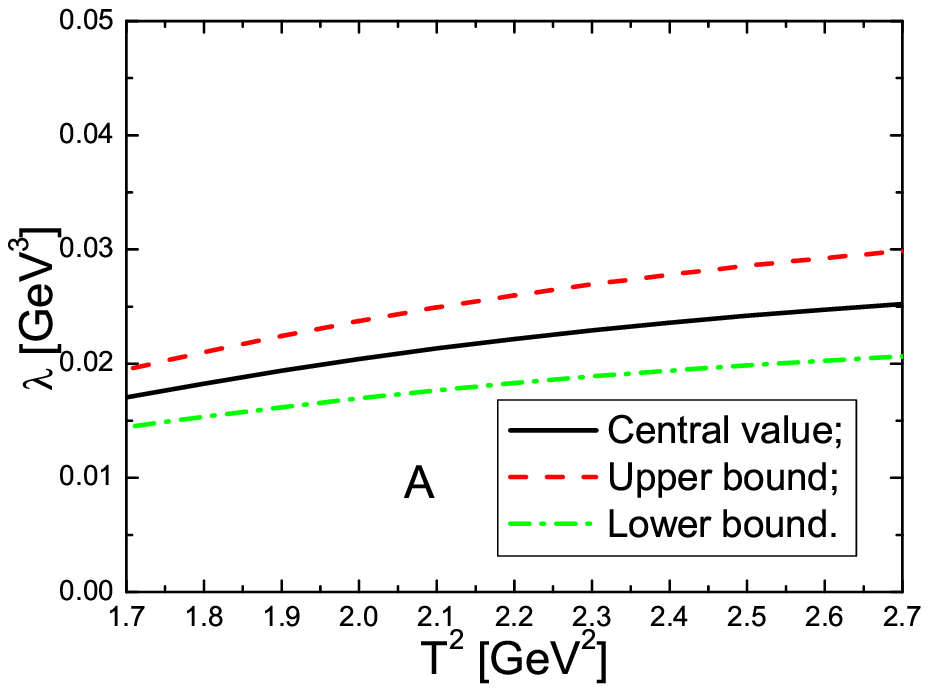}
  \includegraphics[totalheight=4cm,width=5cm]{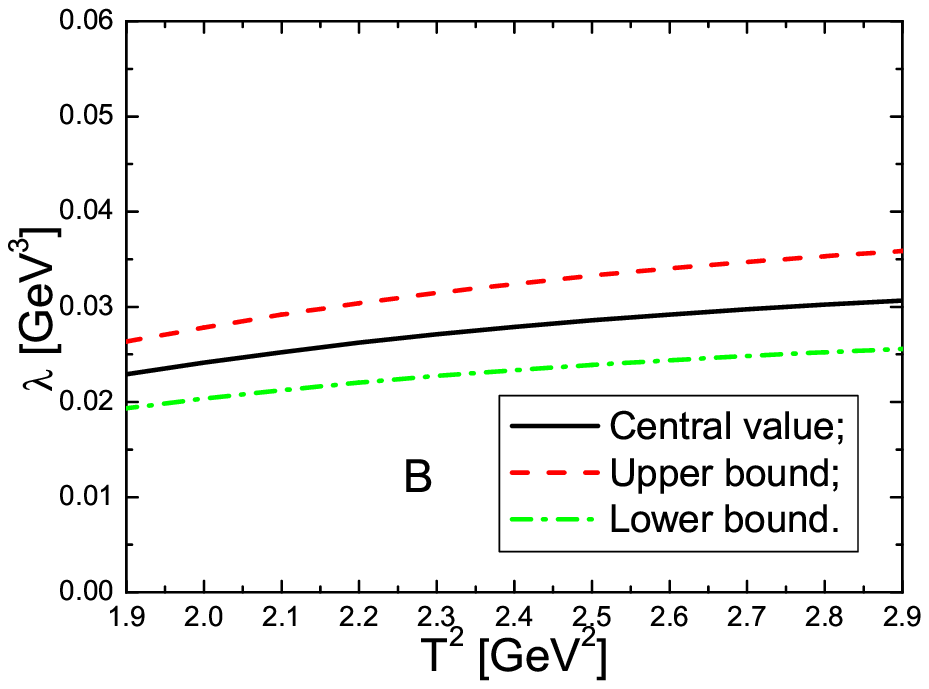}
  \includegraphics[totalheight=4cm,width=5cm]{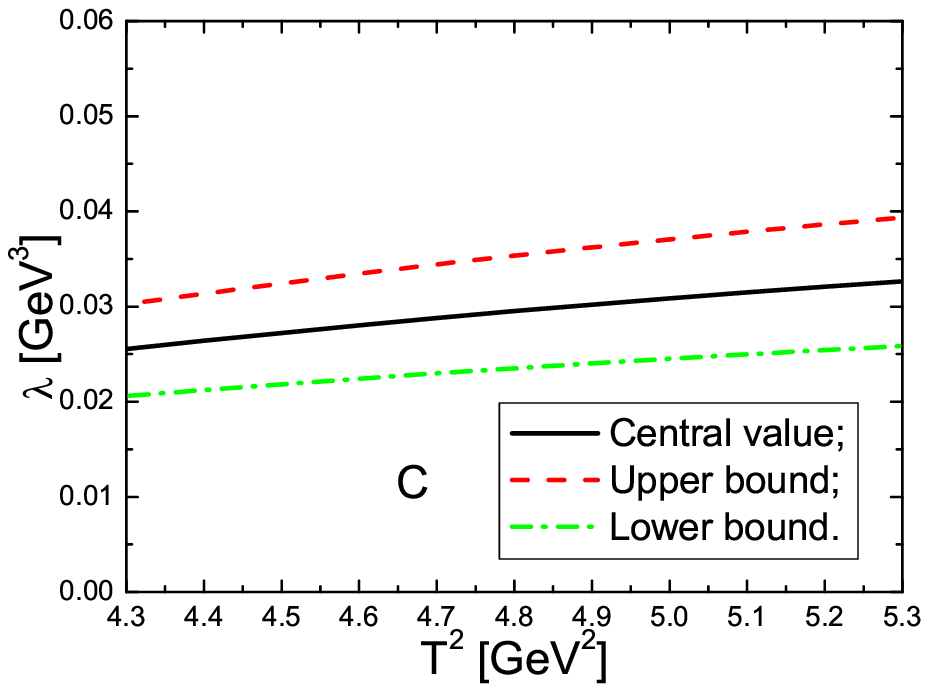}
  \includegraphics[totalheight=4cm,width=5cm]{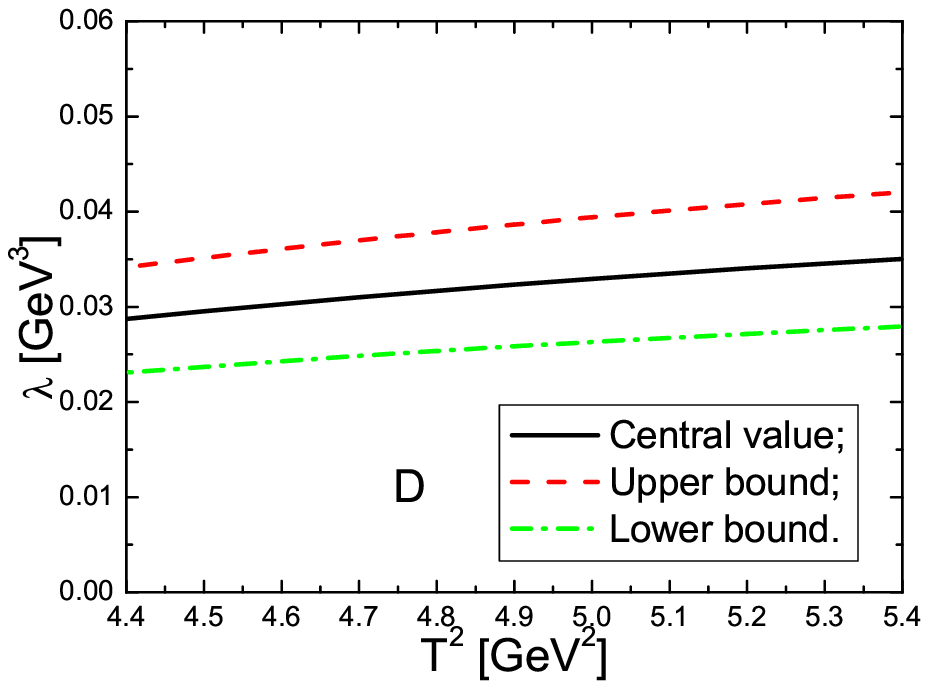}
\includegraphics[totalheight=4cm,width=5cm]{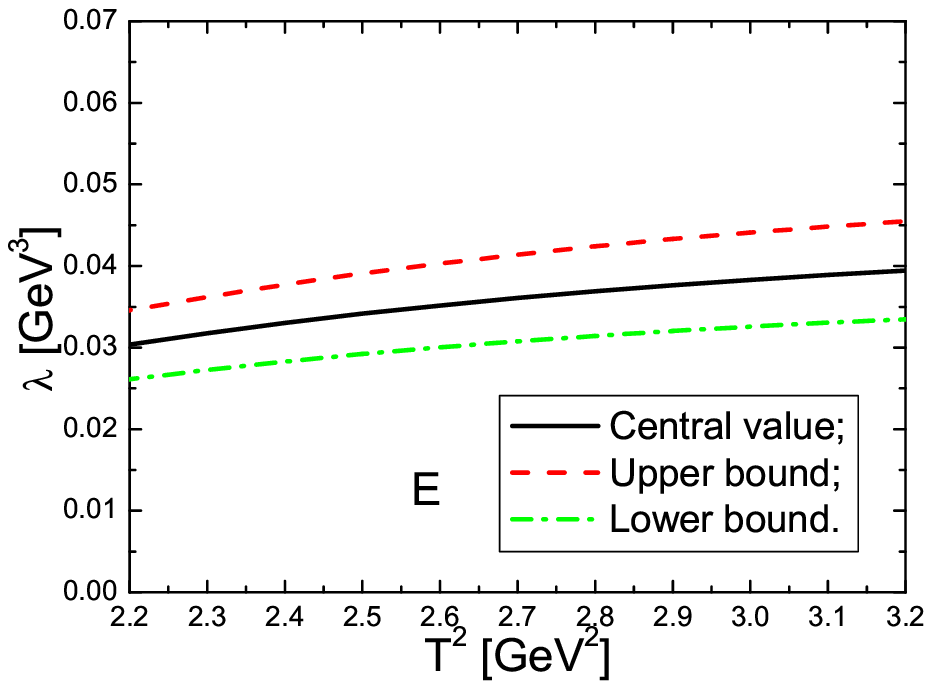}
\includegraphics[totalheight=4cm,width=5cm]{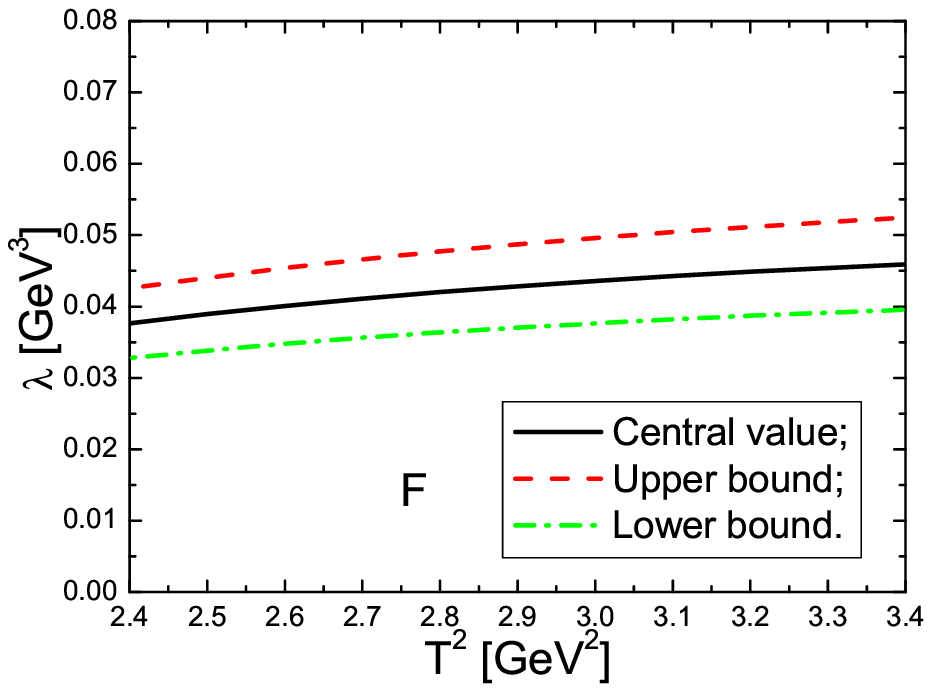}
\includegraphics[totalheight=4cm,width=5cm]{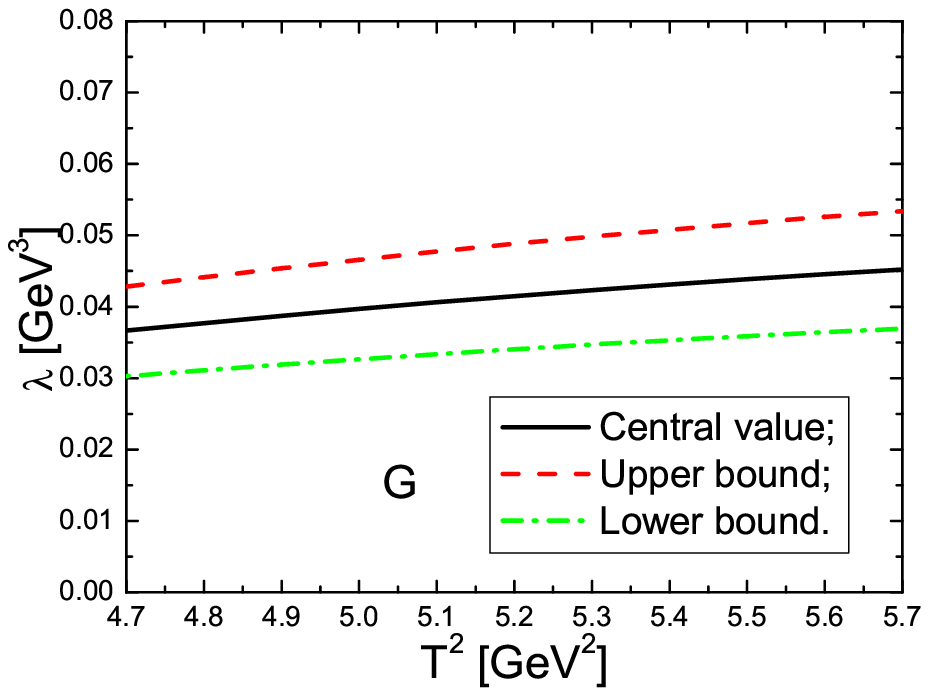}
 \includegraphics[totalheight=4cm,width=5cm]{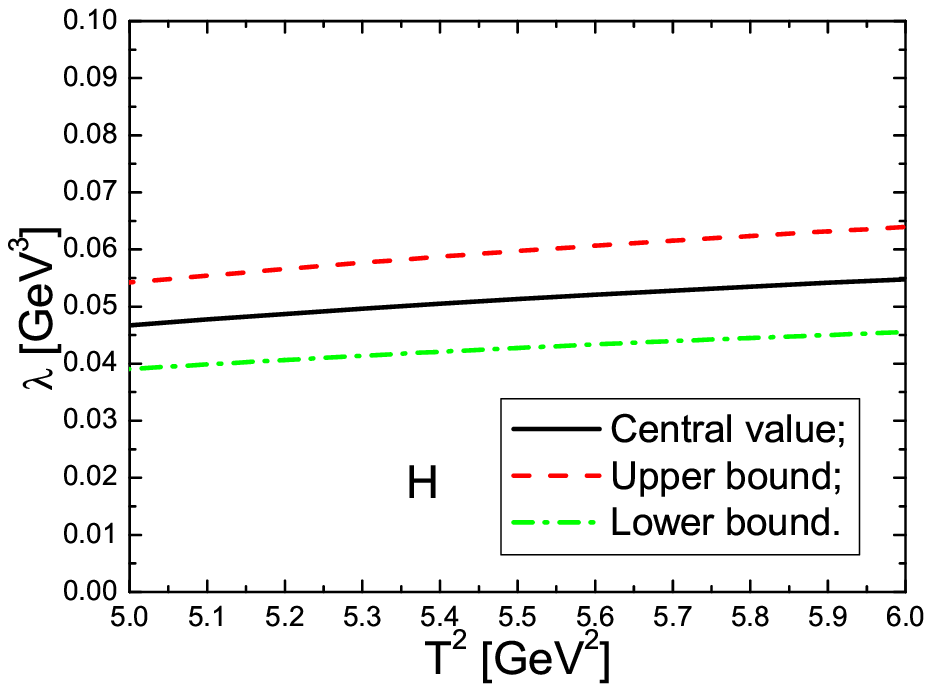}
        \caption{ The pole residues  $\lambda$ of the heavy baryon states, the $A$, $B$, $C$ and $D$  correspond
       to the ${\frac{1}{2}}^+$ heavy baryon channels $\Lambda_c$, $\Xi_c$, $\Lambda_b$ and $\Xi_b$ respectively, while
the $E$, $F$, $G$ and $H$  correspond   to the ${\frac{1}{2}}^-$
heavy baryon channels $\Lambda_c$, $\Xi_c$, $\Lambda_b$ and $\Xi_b$
respectively. }
\end{figure}

In Ref.\cite{Bagan922}, Bagan et al study  the masses and couplings
(pole residues) of the heavy baryon states $\Sigma_b^*$,
$\Sigma_c^*$, $\Lambda_b$ and $\Lambda_c$ with the full QCD sum
rules, and observe that  the $\Sigma_{b(c)}^*$ and $\Sigma_{b(c)}$
have degenerate masses within the uncertainties, while the
$\Lambda_{b(c)}$ is lighter than the $\Sigma_{b(c)}$. In
Refs.\cite{Huang0805,M-Huang0811}, Zhang et al perform a systematic
study of the masses of charmed  and bottom baryon states with  the
full QCD sum rules, where the vacuum condensates  up to dimension
six are taken into account; while in Ref.\cite{M-Nielsen07}, Duraes
et al study  only the masses of the heavy baryon states $\Xi_c$,
$\Xi_b$ and $\Omega_b$. In Ref.\cite{M-Liu07}, Liu et al perform a
systematic study of the masses of the bottom baryon states up to
$\frac{1}{m_Q}$ in the  heavy quark effective field theory using the
QCD sum rules.

In the heavy quark limit, $M_Q=m_Q+\bar{\Lambda}+\mathcal
{O}\left(\frac{1}{m_Q}\right)$. We can calculate the bound energy
$\bar{\Lambda}$ and the $\frac{1}{m_Q}$ corrections of the lowest
heavy baryon states $\Lambda_Q$ with the QCD sum rules,  compare
them with the experimental data $M_Q$, and determine the heavy quark
masses $m_Q$, which always suffer from large uncertainties, then use
the $m_Q$ as basic input parameters to calculate other heavy baryon
masses, one can consult Refs.\cite{Dai961,Dai962,Huang02} for
example.  In Table 3, we also present the predictions from the full
QCD sum rules and the QCD sum rules in the heavy quark effective
theory (where systematic studies are preformed)
\cite{M-Nielsen07,M-Huang0811,M-Liu07}.

From Tables 2-3, we can see that the present and other theoretical
predictions are all in good agreement with the experimental data for
the positive parity baryon states. The negative parity baryon states
$\Lambda_b$ and $\Xi_b$ are not observed yet, we make reasonable
predictions for their masses to confront with experimental data in
the future at the LHCb \cite{LHC}.

The fractions
\begin{eqnarray}
 R&=& \frac{\int_{\Delta}^{\sqrt{s_0}}dp_0\left[\rho^A(p_0) -\rho^B(p_0)\right]\exp\left[-\frac{p_0^2}{T^2}\right]}
{\int_{\Delta}^{\sqrt{s_0}}dp_0\left[\rho^A(p_0)
+\rho^B(p_0)\right]\exp\left[-\frac{p_0^2}{T^2}\right]}
\end{eqnarray}
are less than $4.5\%\,(6.0\%)$ and $0.6\%\,(0.8\%)$ in the positive
(negative) parity charmed  baryon and bottom baryon channels,
respectively. So the contaminations from the negative (or positive)
parity baryon states are very small. In
Refs.\cite{Wang0809,Wang0910}, we study the ${\frac{1}{2}}^+$ flavor
sextet heavy baryon states $\Omega_Q$, $\Xi'_Q$ and $\Sigma_Q$
 with the full QCD sum rules, and observe that  the pole residues
from the sum rules with different tensor structures ($\!\not\!{p}$
and $1$) differ from each other greatly. In Ref.\cite{Wang0912}, we
re-study the masses and pole residues of the ${\frac{1}{2}}^+$
flavor sextet heavy baryon states $\Omega_Q$, $\Xi'_Q$ and
$\Sigma_Q$ by subtracting the contributions of  the negative parity
heavy baryon states, and find that the predictions for the masses
and pole residues are improved considerably. In the present case, we
can choose the tensor structures $\!\not\!{p}$ or $1$ or
$\gamma_0+1$ freely to study the masses and pole residues.

 The pole residues of the $\frac{1}{2}^+$
and $\frac{3}{2}^+$ heavy baryon sextets $B_6$ and $B^*_6$ have been
calculated in our previous works
\cite{Wang0704,Wang0910,Wang0912,Wang1002}. Once reasonable values
of the pole residues $\lambda_{\Lambda}$ and $\lambda_{\Xi}$ of the
$\frac{1}{2}^+$ heavy baryon antitriplet $B_{\bar{3}}$ are obtained,
we can take them as   basic input parameters and study the strong
decays $\Sigma^*_Q\rightarrow\Lambda_Q\pi$,
$\Sigma_Q\rightarrow\Lambda_Q\pi$ and $\Xi^*_Q\rightarrow \Xi_Q \pi$
and the radiative decays $B^*_6 \to B_{\bar{3}}\gamma$ and $B_6 \to
B_{\bar{3}}\gamma$ in a systematic ways with the light-cone QCD sum
rules or the QCD sum rules in external field, and confront  the
predictions with the experimental data in the future at the BESIII,
$\rm{\bar{P}ANDA}$ and LHCb \cite{LHC,BESIII,PANDA}.

 The strong
decays $\Sigma^*_c(2520)\rightarrow\Lambda_c\pi$ and
$\Sigma_c(2455)\rightarrow\Lambda_c\pi$ saturate approximately the
widths of the $\Sigma^*_c(2520)$ and $\Sigma_c (2455)$
 respectively, while the
strong decays $\Xi^*_c(2645)\rightarrow \Xi_c \pi$ are seen
\cite{PDG}. From our previous works \cite{Wang0912,Wang1002}, we can
see that the corresponding strong decays
$\Sigma^*_b\rightarrow\Lambda_b\pi$,
$\Sigma_b\rightarrow\Lambda_b\pi$ and $\Xi^*_b\rightarrow \Xi_b \pi$
are kinematically allowed; although the bottom baryon states
 $\Xi^*_b$ have not been observed experimentally
yet.

In Refs.\cite{Wang0910,Wang0909}, we perform systematic studies  for
the  radiative decays $B^*_6 \to B_{6}\,\gamma$ with the light-cone
QCD sum rules as the strong decays $B^*_6 \to B_{6}\,\pi$ are
forbidden due to the unavailable phase space, while the radiative
channels are not phase space suppressed and become relevant;
although the electromagnetic strength is weaker than that of the
strong interaction. The radiative decays $B^*_6 \to
B_{\bar{3}}\gamma$ and $B_6 \to B_{\bar{3}}\gamma$  are important
processes in testing the heavy quark symmetry and the chiral
symmetry, for example, the $\Xi'_c$ and $\Omega_c^*$ are governed by
the radiative decays.

\section{Conclusion}
In this article, we study the  ${1\over 2}^{\pm}$ flavor antitriplet
heavy baryon states ($\Lambda_c^+$, $\Xi_c^+,\Xi_c^0)$ and
($\Lambda_b^0$, $\Xi_b^0,\Xi_b^-)$ by subtracting the contributions
from the corresponding ${1\over 2}^{\mp}$ heavy baryon states with
the QCD sum rules,  obtain  the masses which  are in good agreement
with the experimental data and make reasonable predictions for the
unobserved ${1\over 2}^-$ bottom baryon states. In calculation, we
observe that the contaminations from the negative (or positive)
parity baryon states are very small, one can choose the tensor
structures $\!\not\!{p}$ or $1$ or $\gamma_0+1$ freely to study the
masses and pole residues. Once reasonable values of the pole
residues $\lambda_{\Lambda}$ and $\lambda_{\Xi}$  are obtained, we
can take them as basic input parameters and study the strong decays
 and radiative decays  in a systematic
ways with the light-cone QCD sum rules or the QCD sum rules in
external field, and confront the predictions  with the experimental
data in the future at the BESIII, $\rm{\bar{P}ANDA}$ and LHCb.

\section*{Acknowledgements}
This  work is supported by National Natural Science Foundation,
Grant Number 10775051, and Program for New Century Excellent Talents
in University, Grant Number NCET-07-0282, and the Fundamental
Research Funds for the Central Universities.


\begin{thebibliography}{99}

\bibitem{ShortRV1} T. Lesiak, hep-ex/0612042.

\bibitem{ShortRV2} J. L. Rosner, J. Phys. Conf. Ser. {\bf 69} (2007) 012002.

\bibitem{ShortRV3}  M. Paulini, arXiv:0906.0808.

\bibitem{SVZ79}  M. A. Shifman, A. I. Vainshtein and V. I. Zakharov,
Nucl. Phys. {\bf B147} (1979) 385, 448.

\bibitem{PRT85} L. J. Reinders, H. Rubinstein and S. Yazaki, Phys. Rept. {\bf
127} (1985) 1.

\bibitem{Bagan93} E. Bagan, M. Chabab, H. G. Dosch and S. Narison, Phys. Lett. {\bf B301}, 243 (1993).

\bibitem{Bagan922} E. Bagan, M. Chabab, H. G. Dosch, and S. Narison,  Phys. Lett.  {\bf B287}, 176 (1992).

\bibitem{Bagan921} E. Bagan, M. Chabab, H. G. Dosch, and S. Narison,  Phys. Lett.  {\bf B278}, 367 (1992).


\bibitem{M-Nielsen07} F. O. Duraes and M. Nielsen, Phys. Lett. {\bf B658} (2007) 40.

\bibitem{Huang0805} J. R. Zhang and M. Q. Huang,  Phys. Rev. {\bf D77} (2008) 094002.

\bibitem{M-Huang0811} J. R. Zhang and M. Q. Huang, Phys. Rev. {\bf D78} (2008) 094015.

\bibitem{Narison0904} M. Albuquerque,  S. Narison and M. Nielsen, Phys. Lett. {\bf B684} (2010) 236.

\bibitem{Shuryak82} E. V. Shuryak, Nucl. Phys. {\bf B198}, 83 (1982).

\bibitem{Grozin92} A. G. Grozin and O. I. Yakovlev, Phys. Lett. {\bf B285}, 254
(1992).


\bibitem{Dai961} Y. B. Dai, C. S. Huang, C. Liu and C. D. Lu, Phys. Lett. {\bf B371}, 99 (1996).

\bibitem{Dai962} Y. B. Dai, C. S. Huang, M. Q. Huang and C. Liu, Phys. Lett. {\bf B387}, 379 (1996).

\bibitem{Huang02} D. W. Wang, M. Q. Huang and C. Z. Li, Phys. Rev.   {\bf D65}, 094036 (2002).

\bibitem{Zhu00} S. L. Zhu, Phys. Rev. {\bf D61}, 114019 (2000).

\bibitem{HuangCS} C. S. Huang, A. L. Zhang and S. L. Zhu, Phys. Lett.  {\bf B492}, 288 (2000).

\bibitem{HuangMQ} D. W. Wang and M. Q. Huang, Phys. Rev. {\bf D68}, 034019(2003).

\bibitem{M-Liu07} X. Liu, H. X. Chen, Y. R. Liu, A. Hosaka and S. L.
Zhu,  Phys. Rev.  {\bf D77}, 014031 (2008).

\bibitem{Wang0704} Z. G. Wang, Eur. Phys. J. {\bf C54} (2008) 231.

\bibitem{Wang0809} Z. G. Wang,  Eur. Phys. J. {\bf C61} (2009) 321.

\bibitem{Wang0910} Z. G. Wang, Eur. Phys. J. {\bf A44} (2010) 105.

\bibitem{Wang0909} Z. G. Wang,  Phys. Rev. {\bf D81} (2010)  036002.

\bibitem{Oka96} D. Jido, N. Kodama and M. Oka,  Phys. Rev. {\bf D54} (1996) 4532.

\bibitem{Chung82} Y. Chung, H. G. Dosch, M. Kremer and D. Schall,  Nucl. Phys. {\bf B197} (1982) 55.

\bibitem{Wang0912} Z. G. Wang, Phys. Lett. {\bf B685} (2010) 59.

\bibitem{Ioffe2005} B. L. Ioffe, Prog. Part. Nucl. Phys. {\bf 56} (2006) 232.

\bibitem{LCSRreview}  P. Colangelo and A. Khodjamirian, hep-ph/0010175.

\bibitem{PDG} C. Amsler et al, Phys. Lett. {\bf  B667} (2008) 1.

\bibitem{NarisonBook} S. Narison, Camb. Monogr. Part. Phys. Nucl. Phys. Cosmol. {\bf 17} (2002) 1.

\bibitem{LHC}  G. Kane and A. Pierce, "Perspectives On LHC Physics",
World Scientific Publishing Company,  2008.


\bibitem{BESIII} D. M. Asner et al, arXiv:0809.1869.

\bibitem{PANDA} M. F. M. Lutz et al, arXiv:0903.3905.


\bibitem{Wang1002} Z. G. Wang,  arXiv:1002.2471.

\end{thebibliography}
\end{document}